\documentclass[preprint, aps,amsmath,amssymb,floats,12pt]{revtex4}
\usepackage{amsmath}
\usepackage{graphicx}
\usepackage{dcolumn}
\usepackage{bm}
\usepackage{lipsum}
\usepackage{pdfpages}

\newcommand{\asinh}{\mathrm{arsinh} \,}

\begin{document}

\title[R Bolle {\em et al}]{An Exposition on the Kaniadakis $\kappa$-Deformed Decay Differential Equation}

\author{Rohan Bolle,  Ibrahim Jarra and Jeffery A. Secrest}
\affiliation{Department of Biochemistry, Chemistry, and Physics and Astronomy\\ 
Georgia Southern University Savannah, GA 31419, USA }
\email{jsecrest@georgiasouthern.edu}

\begin{abstract}
Kaniadakis deformed $\kappa$-mathematics is an area of mathematics that has found relevance in the analysis of complex systems. Specifically, the mathematical framework in the context of a first-order decay $\kappa$-differential equation is investigated, facilitating an in-depth examination of the $\kappa$-mathematical structure. This framework serves as a foundational platform, representing the simplest non-trivial setting for such inquiries which are demonstrated for the first time in the literature. Finally, additional avenues of study are discussed.  
\end{abstract}

%

\maketitle
%
%
%
%

\section{Introduction}
\begin{figure}[t]
\includegraphics[width=0.45\linewidth]{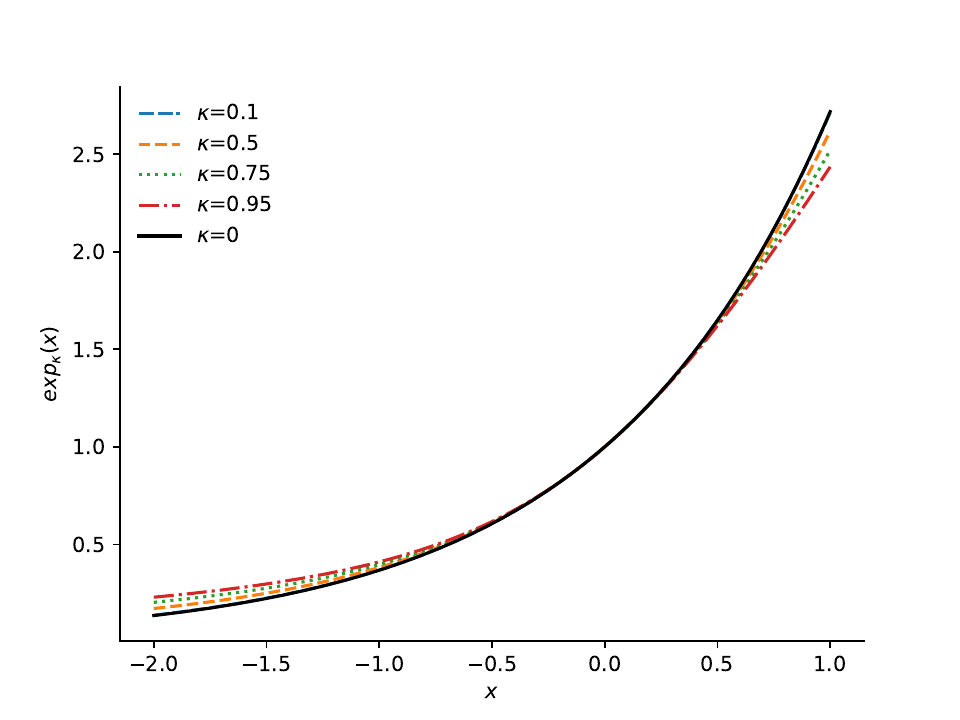}
\includegraphics[width=0.45\linewidth]{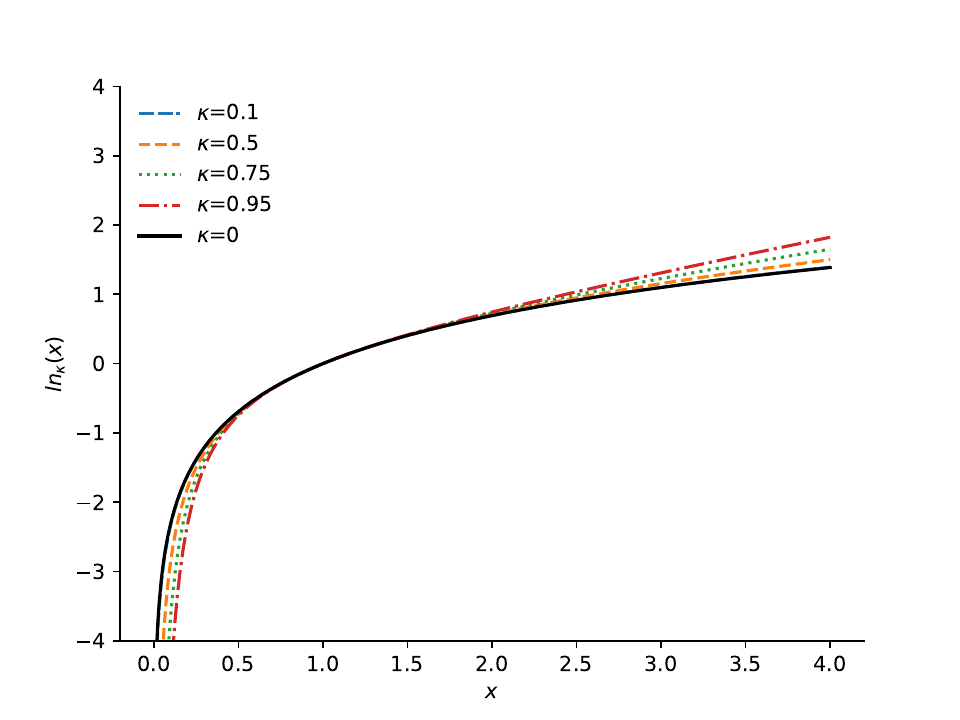}
\caption{Examples of $\kappa$-exponential (left) and $\kappa$-logarithm (right) for various values of deformation parameter $\kappa$.  Note that when $\kappa \rightarrow 0$ then $\kappa$-exponential and $\kappa$-logarithm approach the standard exponential and logarithm.}
\label{fig1}
\centering
\end{figure}
In recent decades the deformed statistics of Kaniadkais  \cite{LandL,Kaniadakis2001, Kaniadakis2002, Kaniadakis2005},
has gained popularity as they have been applied to a wide array of disciplines such as cosmology \cite{cosmo1, cosmo2, cosmo3, cosmo4}, economics \cite{econ1, econ2, econ3}, and epidemiology \cite{Kaniadakis2020} to cite but a few references.  These so-called $\kappa-$statistics are predicated upon the one-parameter of a relativistically inspired generalization of the standard exponential as $\kappa-$exponential,  
\begin{eqnarray}
    \label{eqn:kappa_exp}
    \exp_{\kappa} (x) &=& (\sqrt{1+\kappa^2 x^2} + \kappa x)^{1/\kappa} 
\end{eqnarray}
and the one-parameter relativistically inspired generalization of the usual logarithm as the $\kappa-$logarithm,
\begin{equation}
    \label{eqn:kappa_ln}
    \ln_{\kappa} x = \frac{x^{\kappa} - x^{-\kappa}}{2\kappa}
\end{equation}
where $\kappa$ is a deformation parameter (sometimes also known as the entropic index) taking on values between $-1$ and $1$. Arguably, the most interesting aspect of the $\kappa-$exponential is that it exhibits an exponential behavior at small values of $x$, 
\begin{eqnarray}
 \lim_{x\to0}\exp_{\kappa}(x)\sim \exp(x) 
\end{eqnarray}
whilst exhibiting a power law behavior at asymptotic values of $x$ (that is the tail exhibits a power law structure),
\begin{eqnarray}
 \lim_{x\to\pm\infty}\exp_{\kappa}(x)\sim |2\kappa x|^{\pm 1/|\kappa|}.
\end{eqnarray}
\begin{figure}[t]
\centering
\includegraphics[width=12cm]{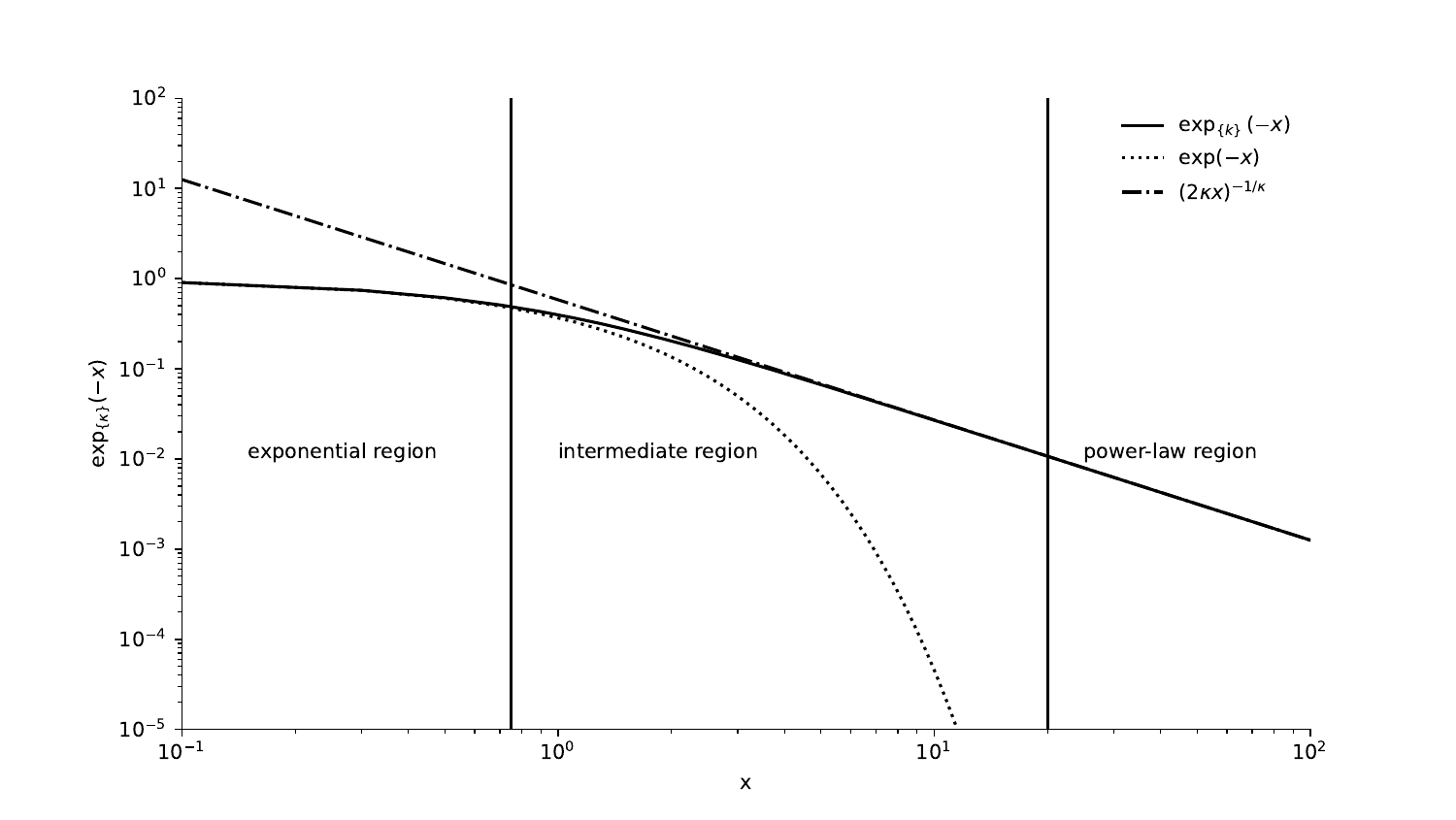}
\caption{The figure above shows the various regions of the $\kappa$-exponential (solid curve), the exponential (dotted curve), and a power-law (dash-dot curve).  The $\kappa$ parameter has the value of 0.75 in the above plots.}
\label{fig_regions}
\end{figure}
Similarly, the $\kappa-$logarithm is similar to the usual standard logarithm as small values of $x$,
\begin{eqnarray}
 \lim_{x\to0}\ln_{\kappa}(x)\sim \ln(x)
\end{eqnarray}
as it also exhibits a power law tails for the asymptotic values of $x$,
\begin{eqnarray}
 \lim_{x\to0^{+}}\ln_{\kappa}(x)\sim\frac{1}{|\kappa|}{x}^{-|\kappa|}\\
 \lim_{x\to+\infty}\ln_{\kappa}(x)\sim \frac{1}{2|\kappa|}{x}^{|\kappa|}.
\end{eqnarray}
Probability distributions with power-law tails are ubiquitous in the natural sciences and this makes these functions interesting.  As mentioned above, some examples of phenomena described by the $\kappa-$statistics are the cosmic ray spectrum \cite{Kaniadakis2002} which extends over 13 decades in energy and 33 decades in particle flux with low energy particles obeying an exponential distribution whilst at high energies the spectrum exhibits a power law tail. The impact of loads on buildings and structures is effectively modeled \cite{civil_eng} using $\kappa-$statistics. In this framework, the $\kappa-$exponential distribution accurately represents normal design loading conditions, while the tails of the distribution follow a power-law pattern, reflecting the effects of loads that exceed design values.

As the deformation parameter approaches zero, 
\begin{eqnarray}
 \lim_{\kappa\to 0} \exp_{\kappa}(x) &\rightarrow \exp(x)\\
 \lim_{\kappa\to 0} \ln_{\kappa}(x) &\rightarrow \ln(x)
\end{eqnarray}
the $\kappa-$exponential and $\kappa-$logarithm approach the standard exponential and natural logarithm respectively. The $\kappa-$exponential and $\kappa-$logarithm are inverses of one another,
\begin{equation}
    \exp_{\kappa}(\ln_{\kappa}(x))=\ln_{\kappa}(\exp_{\kappa}(x))=x.
\end{equation}

The natural question to ask within the framework of these deformed statistics is what is the physical meaning of the $\kappa$-deformation parameter? This deformation parameter has been introduced to describe deviations from the standard exponential behavior, especially in various probability distributions such as the exponential, gaussian, weibull, and pareto distributions. In complicated physical systems, these deformations tend to be related to processes that comprise of short-range interactions, various correlations, or long-term memory effects.  These deformations furthermore describe systems that are not in equilibrium or have a complex hierarchical nature, such as fractal systems.

\section{Overview of $\kappa-$mathematics}
There are two identities that are useful when working with $\kappa-$mathematics,
\begin{equation}
    \asinh =\ln(\sqrt{1+\kappa^2x^2}+\kappa x)\\
\end{equation}
and
\begin{equation}
     \ln x = \asinh\Big[\frac{1}{2}\Big(x-\frac{1}{x}\Big)\Big].
\end{equation}

The algebraic structure $(\mathbb{R},\oplus)$ of the $\kappa$-algebra for $x,y \in \mathbb{R}$ with $\kappa \in [0,1)$ whose $\kappa$-sum composition law is defined via
\begin{equation*}
x \oplus y = x\sqrt{1+\kappa^2y^2} + y\sqrt{1+\kappa^2x^2} = \frac{1}{\kappa}\sinh\Big(\asinh(\kappa x)+\asinh(\kappa y) \Big)
\end{equation*}
forms an abelian group.  The algebraic structure $(\mathbb{R},\otimes)$ for the $x,y$ and $\kappa$ previously described has a $\kappa$-product composition law defined via
\begin{equation*}
x \otimes y = \frac{1}{\kappa}\sinh\Big(\asinh(\kappa x)\asinh(\kappa y) \Big)
\end{equation*}
also forms an abelian group.  Often it is useful to introduce the definition of the $\kappa$-number ${x}_{\kappa} \in C^{\infty}(\mathbb{R})$ via
\begin{equation}
\label{eqn_k_num1}
{x}_{\{\kappa\}} = \frac{1}{\kappa}\asinh(\kappa x)
\end{equation}
and its dual is given as,
\begin{equation}
\label{eqn_k_num2}
{x}^{\{\kappa\}} = \frac{1}{\kappa}\sinh(\kappa x).
\end{equation}

It is easy to show that the $\kappa$-differential is given as
\begin{equation*}
    dx_{\{\kappa \}} = \frac{dx}{\sqrt{1+\kappa^2x^2}}
\end{equation*}
and the $\kappa$-integral which corresponds to the anti-derivative dual,
\begin{equation*}
    \int f(x) dx_{\{\kappa \}} = \int f(x) \frac{dx}{\sqrt{1+\kappa^2x^2}}.
\end{equation*}

The $\kappa$-exponential (see figure \ref{fig1}) is defined as
\begin{eqnarray}
    \label{eqn:kappa_exp}
    \exp_{\kappa} (x) &=& (\sqrt{1+\kappa^2 x^2} + \kappa x)^{1/\kappa} \\ 
                    &=& \exp \Big[ \frac{1}{\kappa}\asinh(\kappa x)\Big] \\
                    &=& \exp({x_{\{\kappa\}}})
\end{eqnarray}
where the Taylor series expansion of the $\kappa$-exponential is determined to be
\begin{eqnarray}
\label{eqn:tay_exp_k}
    \exp_{\kappa}(x) = 1 + &x + \frac{1}{2!}x^2 + (1-\kappa^2)\frac{1}{3!}x^3 + (1-4\kappa^2)\frac{1}{4!}x^4 \nonumber \\
    &+ (1-\kappa^2)(1-9\kappa^2)\frac{1}{5!}x^5 +...
\end{eqnarray}
with the $\kappa$-logarithm (see figure \ref{fig1}) is defined as 
\begin{equation}
    \label{eqn:kappa_ln}
    \ln_{\kappa} x = \frac{x^{\kappa} - x^{-\kappa}}{2\kappa} =\frac{1}{\kappa}\sinh (\kappa \ln x)
\end{equation}
where the Taylor series expansion of the $\kappa$-logarithm is determined to be
\begin{eqnarray}
    \ln_{1+\kappa}(x) = x - &\frac{1}{2}x^2 + \Big(1+\frac{\kappa^2}{2}\Big)\frac{1}{3}x^3-\Big(1+\kappa^2\Big)\frac{1}{4}x^4 \nonumber \\
    &+\Big(24+35\kappa^2+\kappa^4\Big)\frac{1}{5!}x^5+...
\end{eqnarray}
where the deformation parameter $\kappa$ exists on the interval, $|\kappa|<1$ and as $\kappa \rightarrow 0$ equations~(\ref{eqn:kappa_exp}) and (\ref{eqn:kappa_ln})  reduces to the usual logarithm and exponential functions. Equations~(\ref{eqn:kappa_exp}) and (\ref{eqn:kappa_ln}) are inverses of each other, and thus the following property holds, 
\begin{eqnarray}
\ln_{\kappa}(\exp_{\kappa} x)=\exp_{\kappa}(\ln_{\kappa} x)=x. \nonumber
\end{eqnarray}

\begin{figure}[t]
\centering
\includegraphics[width=8cm]{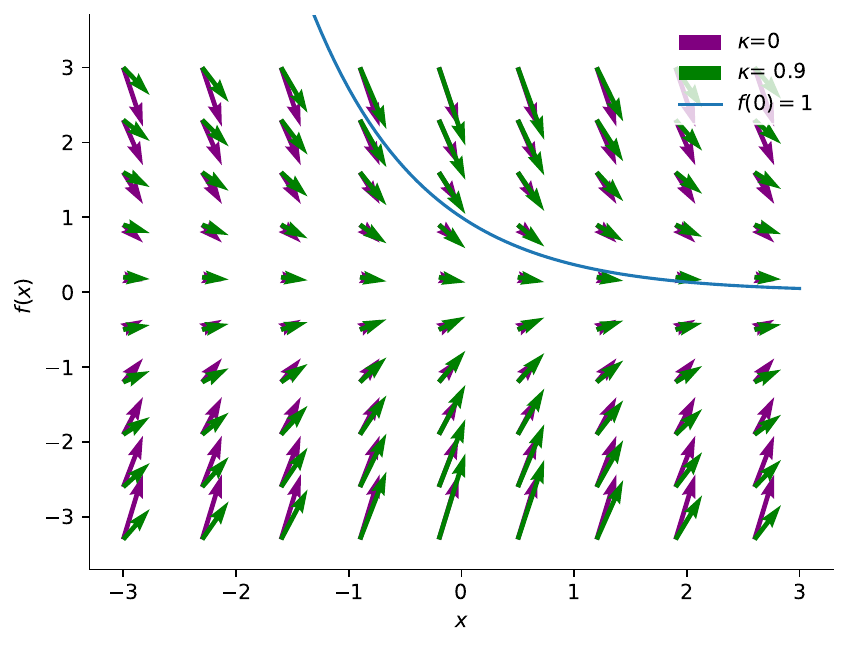}
\caption{The slope field for $\kappa$-decay differential equation (\ref{eqn:kappa_ode}) for $\kappa=0.9$ and for $\kappa \rightarrow 0$ which leads to standard decay differential equation (\ref{eqn:ode}).  The blue curve represents the solution, $f(x)=\exp_{\kappa}(-x)$, with the initial condition that $f(0)=1$.}
\label{fig2}
\end{figure}

One of the most ubiquitous equations in physics is the linear first-order homogeneous decay differential equation \cite{decay_exp_ode, arfken, boas},
\begin{equation}
    \frac{df(x)}{dx}+r(x)f(x)=0
    \label{eqn:ode}
\end{equation}
where in this work the rate function $r(x)=r_0=1$ is taken to be constant and the initial condition $f(0)=1$.  In this particular instance, the rate of decrease is a function, $f(x)$, that is proportional to the value of $f(x)$ itself which leads to the well-known standard decaying exponential function,
\begin{equation}
\label{eqn:std_exp}
    f(x) = \exp(-x).
\end{equation}
This equation describes radioactive half-lives, simple population decay, discharge of a capacitor, and rate law for certain first-order molecular reactions.  In this work, the homogeneous first-order $\kappa$-differential equation, 
\begin{equation}
    \sqrt{1+\kappa^2x^2}\frac{df(x)}{dx}+r(x)f(x)=0
    \label{eqn:kappa_ode}
\end{equation}
will be studied where again the rate function $r(x)=r_0=1$ is taken to be a constant and the initial condition $f(0)=1$. The solution to this equation is analogous to the solution to equation (\ref{eqn:std_exp}),
\begin{equation}
    f(x) = \exp_{\kappa}(-x).
\end{equation}

\section{Techniques}
In order to gain a deeper understanding of the $\kappa$-deformed framework, the deformed differential equation described in equation (\ref{eqn:kappa_ode}) is solved by a variety of elementary methods to demonstrate the mathematical structure of the differential equation as well as the solution.

\begin{enumerate}
        \item Validation and verification: validating and verifying solutions in the context of differential equations are important steps in the process of solving and understanding these equations and their solutions. These two steps ensure that the obtained solutions are accurate and consistent with the original differential equation and any initial or boundary conditions.
        \item Understanding: the $\kappa$-deformed statistics are not as well known as standard statistics or other deformed statistics (such as Tsallis statistics).  This is an opportunity to investigate these statistics and in particular the deformed exponential and the deformed decay equation from which it is a solution within a familiar framework.  By exploring a variety of methods for solving the same differential equation, valuable experience and insights that contribute to a deeper understanding of both the equation itself and its solutions can be acquired.
        \item Rigor: Solving a problem through various methods introduces rigor by subjecting the solution to different analytical approaches, verification steps, and perspectives. This multiplicity of approaches helps ensure the robustness and reliability of the solution. It allows for cross-validation and comparison
        \item Computational efficiency and Applicability: this presents an opportunity to examine this deformed differential equation and corresponding solution and to determine which methods are better for handling the additional complexities introduced by the deformation aspects of both the equation and solutions.
\end{enumerate}

\subsection{Direct Substitution}
The method of Direct Substitution is a technique used to solve certain types of ordinary differential equations. This method involves substituting a proposed solution into the differential equation to check if it satisfies the differential equation. If the substituted solution works, it is the solution to the differential equation. This method is advantageous when wanting to confirm a guessed and given solution. Assuming the solution has the form of equation (\ref{eqn:kappa_exp}) and taking the derivative yields, 
\begin{eqnarray*}
    \frac{df(x)}{dx} &=& \frac{d}{dx}\exp_{\kappa}(-x)\\
     &=& \frac{d}{dx} \Big( \sqrt{1+\kappa^2x^2} - \kappa x \Big)^{1/\kappa}\\
                                  &=&\frac{-1}{\sqrt{1+\kappa^2x^2}}\exp_{\kappa}(-x) \\
                                  &=&\frac{-1}{\sqrt{1+\kappa^2x^2}}f(x)\\
\end{eqnarray*}
and rearranging leads to equation (\ref{eqn:kappa_ode}).  This method provides an intuitive way to understand the relationship between the differential equation and the $\kappa-$exponential.

\subsection{Separation of Variables I}
\label{sec:sovI}
The method of Separation of Variables is one of the most straightforward techniques to determine the solution to a differential equation.  This method requires that the differential equation be written in the form,
\begin{eqnarray*}
    m(x)+n(f)\frac{df}{dx}=0
\end{eqnarray*}
which can then be manipulated into the form,
\begin{equation*}
\label{eqn:sep}
    \int n(f)df = -\int m(x)dx
\end{equation*}
where direct integration can be applied in this situation.

The $\kappa$-decay equation (\ref{eqn:kappa_ode}) can be separated into the form of equation (\ref{eqn:sep}) as,
\begin{equation*}
    \int_{1}^{f}\frac{df(x)}{f(x)} = -\int_{0}^{x}\frac{1}{\sqrt{1+\kappa^2 x^2}} dx 
\end{equation*}
which leads to 
\begin{eqnarray}
    f(x) &=& \exp\left[-\int_{0}^{x}\frac{1}{\sqrt{1+\kappa^2x^2}}dx\right] \nonumber \\
         &=& \exp\left[-\frac{\asinh(\kappa x)}{\kappa}\right] \nonumber \\ 
         &=& \exp_{\kappa}(-x). \nonumber
\end{eqnarray}

\subsection{Separation of Variables II}
Another more interesting application of the method of Separation of Variables that was described in Section \ref{sec:sovI} can be applied to the $\kappa$-decay equation utilizing the $\kappa$-numbers described by equations (\ref{eqn_k_num1}) and  (\ref{eqn_k_num2}).

Rewriting equation (\ref{eqn:ode}),
\begin{eqnarray}
   \frac{df(x_{\{\kappa\}})}{dx_{\{\kappa\}}} = -f(x_{\{\kappa\}})  \nonumber
\end{eqnarray}
and separating,
\begin{eqnarray}
    \int_{1}^{f}\frac{df(x_{\{\kappa\}})}{f(x_{\{\kappa\}})} = -\int_{0}^{x_{\{\kappa\}}}dx_{\{\kappa\}} \nonumber 
\end{eqnarray}
and integrating yields,
\begin{eqnarray}
    \ln f(x_{\{\kappa\}}) = -x_{\{\kappa\}} \nonumber 
\end{eqnarray}
and exponentiating both sides,
\begin{eqnarray}
    f(x_{\{\kappa\}}) &=& \exp(-x_{\{\kappa\}}) \nonumber 
\end{eqnarray}
and applying the relationship described by equation (\ref{eqn:kappa_exp}) results in 
\begin{eqnarray}
                      &=& \exp_{\kappa}(-x). \nonumber
\end{eqnarray}

\subsection{Integrating Factor}
The technique of the Integrating Factor \cite{arfken, boas, IF1, IF2} is a special implementation of the separation of variables technique with the derivative product rule.  It should be noted that the integrating factor can be identified as the Green's function \cite{IF3} when dealing with a non-homogenous first-order differential equation. This method requires the first-order differential equation of interest to be of the form,
\begin{eqnarray}
    \frac{df}{dx} +P(x)f=Q(x)
    \label{eqn:gen_1st_ode}
\end{eqnarray}
which almost resembles the product rule from elementary calculus. Multiplying both sides of equation (\ref{eqn:gen_1st_ode}) by $I(x)$ results in, 
\begin{eqnarray*}
    I(x)\frac{df}{dx} +I(x)P(x)f=I(x)Q(x)
\end{eqnarray*}
where this looks somewhat like the derivative of $I(x)f(x)$.  In order to be a separable equation it requires that,
\begin{eqnarray*}
    \frac{dI}{dx} = I(x)P(x) 
\end{eqnarray*}
which leads to the separable equation, 
\begin{eqnarray*}
\label{eqn:if_eqn}
    \frac{df}{dx} = f(x)P(x) 
\end{eqnarray*}
where the solution is, 
\begin{eqnarray}
\label{eqn:if_soln}
    f(x)=A\exp[I] 
\end{eqnarray}
and the integration factor, $I(x)$, 
\begin{eqnarray}
\label{eqn:if}
    I=-\int P(x)dx.
\end{eqnarray}

Examining the $\kappa$-decay equation, equation (\ref{eqn:kappa_ode}), the quantities $P(x)=\frac{1}{\sqrt{1+\kappa^2x^2}}$ and $Q(x)=0$ as defined in equation (\ref{eqn:gen_1st_ode}).  Using equations (\ref{eqn:if_soln}) and (\ref{eqn:if})  leads to the result, 
\begin{eqnarray}
    f(x)=A\exp\Big[-\int_{0}^{x} \frac{1}{\sqrt{1+\kappa^2x^2}}\Big] \nonumber
\end{eqnarray}
which when the initial condition, $f(0)=1$ is applied, the constant $A$ is determined, and the solution is,
\begin{eqnarray}
                     f(x) = \exp_{\kappa}(-x). \nonumber
\end{eqnarray}

\subsection{WKBJ}
The Wentzel–Kramers–Brillouin-Jeffreys (WKBJ) method though relatively specialized can be applied to equation (\ref{eqn:ode}) in order to determine the solution \cite{WKB:1981}.  It is assumed the solution will be of the form,
\begin{eqnarray}
   f = A\exp\{B(x)\} \nonumber
   \label{eqn:wkb_test_soln}
\end{eqnarray}
where $A$ is a constant and $B(x)$ satisfies the differential equation
\begin{eqnarray}
   \sqrt{1+\kappa^2x^2}B^{'}=-1 \nonumber
\end{eqnarray}
where
\begin{eqnarray}
    B^{'}=\frac{dB}{dx}
\end{eqnarray}
which leads to 
\begin{eqnarray}
    B &=& \int \frac{dx}{\sqrt{1+\kappa^2x^2}} \nonumber \\
      &=& -\frac{\asinh(\kappa x)}{\kappa}.
      \label{eqn:wkb_phase}
\end{eqnarray}
Plugging the result for the exponential's exponent, equation (\ref{eqn:wkb_phase}) into the test function equation (\ref{eqn:wkb_test_soln}), is
\begin{eqnarray}
   f = A\exp\Big[-\frac{\asinh(\kappa x)}{\kappa}\Big] \nonumber \\ 
     = A \exp_{\kappa}(-x) \nonumber \\ 
     = \exp_{\kappa}(-x) \nonumber 
\end{eqnarray}
where $A = 1$ in order to satisfy the initial condition that $f(0)=1$.

\section{Laplace Transform}
The Laplace transform is a technique to reduce a differential equation to an algebraic equation.  The Laplace transform of the function $g(t)$ is denoted by $\mathcal{L}\{g(t)\}$ and $G(s)$ is defined by
\begin{eqnarray*}
    \mathcal{L}\{g(t)\} = \int_{0}^{\infty}f(t)e^{-st}dt = G(s).
\end{eqnarray*}

Rewriting equation (\ref{eqn:ode}) using the $\kappa$-numbers defined by equations (\ref{eqn_k_num1}) and (\ref{eqn_k_num2}),
\begin{eqnarray}
   \frac{df(x_{\{\kappa\}})}{dx_{\{\kappa\}}} = -f(x_{\{\kappa\}})  \nonumber
\end{eqnarray}
which can be re-written as
\begin{eqnarray}
   f^{\prime}(x_{\{\kappa\}}) + f(x_{\{\kappa\}}) = 0  \nonumber
\end{eqnarray}
where the notation of the expression has been simplified using the $\prime$-notation to describe the derivative with respect to $x_{\{\kappa\}}$.  Applying the Laplace transform, $\mathcal{L}$,
\begin{eqnarray}
  \mathcal{L}\{ f^{\prime}(x_{\{\kappa\}}) + f(x_{\{\kappa\}})\} =  \mathcal{L}\{0\}  \nonumber
\end{eqnarray}
which leads to 
\begin{eqnarray}
  \mathcal{L}\{ f^{\prime}(x_{\{\kappa\}})\} + \mathcal{L}\{f(x_{\{\kappa\}})\} =  0  
  \label{laplace_de}
\end{eqnarray}
where the Laplace transform of a sum is the sum of Laplace transforms and the Laplace transform of zero is simply zero.  Applying the Laplace transform for a derivative,
\begin{eqnarray*}
  \mathcal{L}\{ g(t)^{\prime} \}  = s \mathcal{L}\{ g(t) \} -g(0)   
\end{eqnarray*}
to equation (\ref{laplace_de}) leads to
\begin{eqnarray}
  \mathcal{L}\{f(x_{\{\kappa\}})\}   =   \frac{f(0)}{y+1}
  \label{eqn:laplace_transformed}
\end{eqnarray}
and using the inverse Laplace transform to equation (\ref{eqn:laplace_transformed}) which leads to
\begin{eqnarray*}
  f(x_{\{\kappa\}})  =  f(0)\mathcal{L}^{-1} \Bigl\{   \frac{1}{y+1}  \Bigr\}
\end{eqnarray*}
where the inverse Laplace transform is known to be
\begin{eqnarray*}
  \mathcal{L}^{-1}\Bigl\{ \frac{1}{s+\alpha} \Bigr\} = \exp(-\alpha t)   
\end{eqnarray*}
which leads to the result
\begin{eqnarray*}
  f(x_{\{\kappa\}})  =  f(0)\exp(-x_{\{\kappa\}})
\end{eqnarray*}
and applying the initial condition that $f(0)=1$ leads to 
\begin{eqnarray}
                      \exp(-x_{\{\kappa\}}) = \exp_{\kappa}(-x). \nonumber
\end{eqnarray}
where the relationship described by equation (\ref{eqn:kappa_exp}) was applied and the $\kappa$-exponential is recovered as expected.

\section{Lagrange-Charpit Method}
The Lagrange-Charpit method \cite{L-C1} also known as the Method of Characteristics is a technique for solving hyperbolic differential equations.

Consider the partial differential equation of the form,
\begin{equation}
a(x,y)\frac{\partial f}{\partial x} + b(x,y) \frac{\partial f}{\partial y}=c(x,y,f)     
\label{gen_LCM}
\end{equation}
where if $f(x,y)$ is a solution then at each point $(x,y)$ equation (\ref{gen_LCM}) can be rewritten as 
\begin{equation*}
    (a(x,y),b(x,y),c(x,y))\cdot \Big(\frac{\partial f}{\partial x}, \frac{\partial f}{\partial y},-1\Big)=0
\end{equation*}
which due to the dot product implies that these two vectors are orthogonal at that point.  Consequently, in order to determine the solution on the surface described by a point $(x,y,f(x,y))$, the vector $(a(x,y),b(x,y),c(x,y))$ must line in the tangent plane.  In order to determine the characteristic curve described by the reparameterized vector $(a(\sigma),b(\sigma),c(\sigma))$ satisfies the following system of differential equations,
\begin{eqnarray*}
    \frac{dx}{d\sigma} &=& a(x(\sigma),y(\sigma)),\\
    \frac{dy}{d\sigma} &=& b(x(\sigma),y(\sigma)),
\end{eqnarray*}
and
\begin{eqnarray*}        
    \frac{df}{d\sigma} &=& c(x(\sigma),y(\sigma),f(\sigma))
\end{eqnarray*}
which yields the so-called Lagrange-Charpit equations,
\begin{equation*}
    \frac{dx}{a}=\frac{dy}{b}=\frac{df}{c}.
\end{equation*}
Applying this technique to the $\kappa$-decay differential equation, the coefficients of equation (\ref{gen_LCM}) from equation (\ref{eqn:kappa_ode}) can be read off as
\begin{eqnarray*}
a &=& \sqrt{1+\kappa^2x^2}\\
b &=& 0
\end{eqnarray*}
and
\begin{eqnarray*}    
c &=& -u
\end{eqnarray*}
which leads to the integrals,
\begin{equation*}
        \int_{1}^{f}\frac{df}{f} = -\int_{0}^{x}\frac{dx}{\sqrt{1+\kappa^2x^2}} 
\end{equation*}
which results in the solution,
\begin{eqnarray}
                      = \exp_{\kappa}(-x). \nonumber
\end{eqnarray}

\section{Picard's Iterative Method}
Picard's Method \cite{Picard1, Picard2}, sometimes known as the Method of Successive Approximations, is a technique where one transforms a differential equation of the form, 
\begin{equation*}
\frac{df(x)}{dx} = G(x,f), \hspace{0.25cm} f(x_0)=f_0.
\end{equation*}
into an integral equation and then apply the fundamental theorem of calculus to solve the integral equation 
where the fundamental theorem of calculus can now be applied  to the above equation to yield, 
\begin{equation*}
 f(x)-f(x_0) = \int_{x_0}^x G(x,f)dx 
\end{equation*}
which reduces the differential equation to the equivalent integral equation,
\begin{equation*}
     f(x) = f_0 +\int_{x_0}^{x}G(x,f(x))dx
\end{equation*}
In order to solve this integral equation, which can be written iteratively as, 
\begin{equation*}
     f(x) = f_0 +\int_{x_0}^{x}G(x,f_n(x))dx
\end{equation*}
Picard's Iteration algorithm can be applied:
\begin{enumerate}
    \item choose an initial guess, $n=0$ for $f_0$ where $f_0$ is the initial condition, and 
    \item choose $f_n = \int G(x,f_n(x))dx$.
\end{enumerate}

\begin{eqnarray*}
    f_n = f_0 +\int_{x_0}^{x}G(t,f_{n-1}(t))dt
\end{eqnarray*}
The first approximation is with $f_0 = 1$, then second approximation becomes,
\begin{eqnarray}
    f_1 &=& f_0 + \int_{0}^{x}G(t,y_{0}(t))dt \nonumber \\
        &=& 1-\int_0^x \frac{1}{\sqrt{1+\kappa^2t^2}} \, dt \nonumber \\
        &=& 1-\frac{\asinh(\kappa x)}{\kappa}.
        \label{eqn:picard_approx_1}
\end{eqnarray}
The next approximation using the result given by the result of equation (\ref{eqn:picard_approx_1}) is
\begin{eqnarray*}
    f_2 &=& f_0 + \int_{0}^{x}G(t,y_{1}(t))dt \nonumber \\
        &=& 1-\int_0^x \frac{\left(1-\frac{\asinh(\kappa t)}{\kappa}\right)}{\sqrt{1+\kappa^2t^2}} \, dt \nonumber \\
        &=& \frac{1}{2}+\frac{(\kappa-\asinh(\kappa x))^2}{2 \kappa^2}
        \label{eqn:picard_approx_2}
\end{eqnarray*}
The next approximation using the result given by
\begin{equation*}
    f_3 = 1-\frac{\asinh(\kappa x)\left(6 \kappa^2-3 \kappa \asinh(\kappa x)+\asinh(\kappa x)\right)}{6 \kappa^3} 
\end{equation*}
\begin{eqnarray}
    f_4 = 1&+\frac{\asinh(\kappa x)}{24 \kappa^4} \big(-24 \kappa^3+12 \kappa^2 \asinh(\kappa x)  \nonumber \\
    &- 4 \kappa \asinh(\kappa x)^2+\asinh(\kappa x)^3 \big)   \nonumber
    \end{eqnarray}
\begin{eqnarray}
    f_5 =1&-\frac{\asinh(\kappa x)}{120 \kappa^5}\big(120 \kappa^4 -60 \kappa^3 \asinh(\kappa x) \nonumber \\
    &+20 \kappa^2 \asinh(\kappa x)^2-5 \kappa \asinh(\kappa x)^3+\asinh(\kappa x)^4 \big) 
    \label{eqn:picard_approx_5}
\end{eqnarray}

\begin{figure}[h]
\centering
\includegraphics[width=7cm]{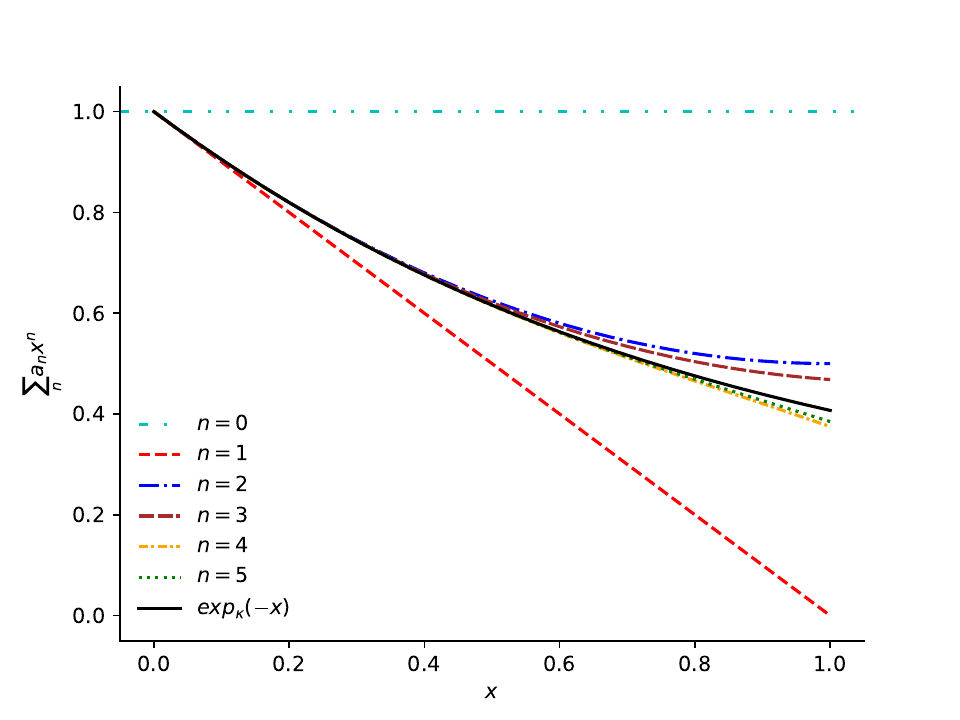}
\includegraphics[width=7cm]{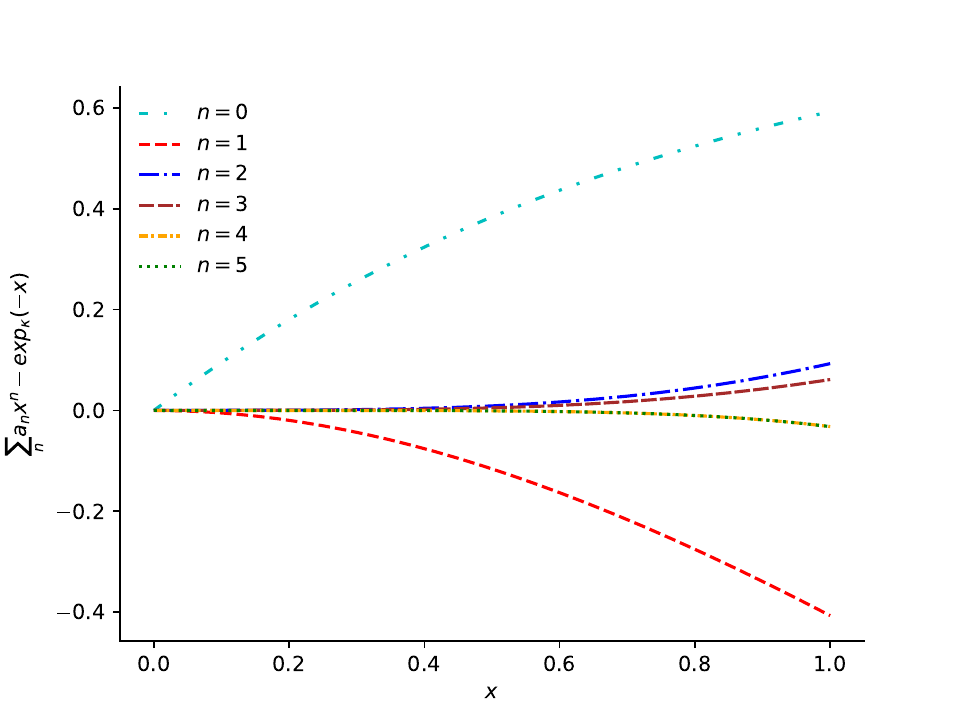}
\caption{The figure on the left is a comparison between the series solution (of various highest orders) from Picard's Method (see equation (\ref{eqn:picard_soln})) and the Power Series Method (see equation (\ref{eqn:series_soln})) and the exact solution, $f(x)=\exp_{\kappa}(-x)$. The figure on the right is the difference between the power series truncated at order $n$ and the $\kappa$-exponential.}
\label{fig3}
\end{figure}

Taylor series expanding the result (\ref{eqn:picard_approx_5}) leads to 
\begin{eqnarray}
\label{eqn:picard_soln}
f(x) \approx 1-x+\frac{x^2}{2}+\frac{1}{6} \left(-1+\kappa^2\right) x^3+\frac{1}{24} \left(1-4 \kappa^2\right) x^4 + \mathcal{O}(x^5)\
\end{eqnarray}
which is in agreement with the Taylor series described by equation (\ref{eqn:tay_exp_k}) which leads to the $\kappa$-decaying exponential,
\begin{eqnarray*}
f(x) =\exp_{\kappa}(-x).
\end{eqnarray*}
The series solution for various highest orders is shown in figure \ref{fig3} along with the associated errors.

\section{Power Series Method}
The Power Series Method seeks to determine a series solution by utilizing a trial power series solution \cite{Series1}.  The trial solution along with its derivative are taken to be
\begin{eqnarray}
    \label{eqn:series_ansatz}
    f(x) &=& \sum_{n=0}^{\infty} a_n x^n = a_0 + a_1 x + a_2 x^2 + a_3 x^3 +a_4 x^4 + a_5 x^5 + \mathcal{O}(x^6)\\
    \frac{df(x)}{dx} &=& \sum_{n=1}^{\infty} = a_1  + 2 a_2 x + 3 a_3 x^2 + 4 a_4 x^3 + 5 a_5 x^4 + \mathcal{O}(x^5)
\end{eqnarray}

Plugging the ansatz into equation (\ref{eqn:gen_1st_ode}) results in
    \begin{eqnarray*}
        \sqrt{1+\kappa^2 x^2}(a_0 + a_1 x + & a_2 x^2 + a_3 x^3 
        +  a_4 x^4  + a_5 x^5 + \mathcal{O}(x^6)) \\
        & = -(a_1  + 2 a_2 x + 3 a_3 x^2 + 4 a_4 x^3 + 5 a_5 x^4 + \mathcal{O}(x^5))
    \end{eqnarray*}
and applying Taylor series to expand the square root,
    \begin{eqnarray*}
        \Big(1+\frac{\kappa ^2}x^2 - \frac{\kappa^4 x^4}{8} + \mathcal{O}(x^6)\Big) & \Big(a_0 + a_1 x + a_2 x^2 + a_3 x^3 
        + a_4 x^4  + a_5 x^5 + \mathcal{O}(x^6)\Big) \\
        &= -\Big(a_1  + 2 a_2 x + 3 a_3 x^2 + 4 a_4 x^3 + 5 a_5 x^4 + \mathcal{O}(x^5)\Big)
    \end{eqnarray*}
 
Equating coefficients of various powers of $x$ leads to the following relationships,
\begin{eqnarray}
    a_1 = -a_0,                              \nonumber \\
    2a_2 = -a_1,                 \nonumber\\
    3a_3 + \frac{\kappa^2}{2}a_1 = -a_2,     \nonumber  \\
    4a_4 + \kappa^2 a_2 = -a_3,               \nonumber 
\end{eqnarray}
and
\begin{eqnarray}    
    5a_5 + 3 \frac{\kappa^2}{2} a_3 - \frac{\kappa^4}{8}a_1 = -a_4.   \nonumber 
\end{eqnarray}
Using the initial condition that $f(0)=1$ allows us to solve for $a_0$ which then allows for the subsequent determination of the other coefficients yielding
\begin{eqnarray}
    a_0 &=& 1,  \nonumber  \\
    a_1 &=& -1,   \nonumber \\
    a_2 &=& \frac{1}{2},   \nonumber  \\
    a_3 &=& \frac{1}{6}(\kappa^2 - 1),    \nonumber  
\end{eqnarray}
and
\begin{eqnarray}        
    a_4 &=& \frac{1}{24}(1-4\kappa^2)  \nonumber 
\end{eqnarray}
plugging these coefficients back into equation (\ref{eqn:series_ansatz}) results in,
\begin{eqnarray}
\label{eqn:series_soln}
f(x) \approx 1-x+\frac{x^2}{2}+\frac{1}{6} \left(\kappa^2-1\right) x^3+\frac{1}{24} \left(1-4 \kappa^2\right) x^4 + \mathcal{O}(x^5)\
\end{eqnarray}
which results in the Taylor series described by equation (\ref{eqn:tay_exp_k}) which leads to the $\kappa$-decaying exponential,
\begin{eqnarray*}
f(x) =\exp_{\kappa}(-x).
\end{eqnarray*}
The series solution for various highest orders is shown in figure \ref{fig3} along with the associated errors.

\section{Numerical Methods}
\begin{figure}[h]
\centering
\includegraphics[width=7cm]{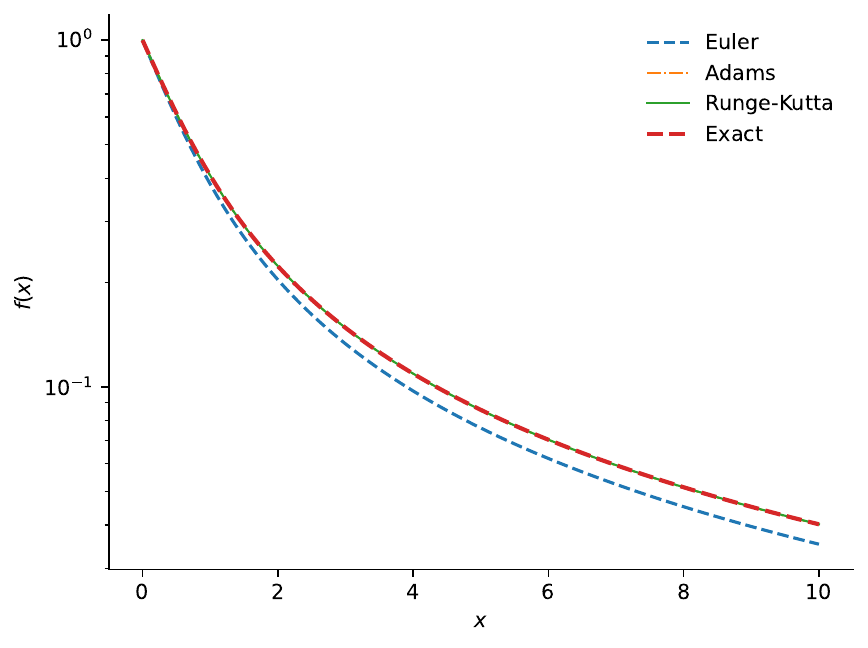}
\includegraphics[width=7cm]{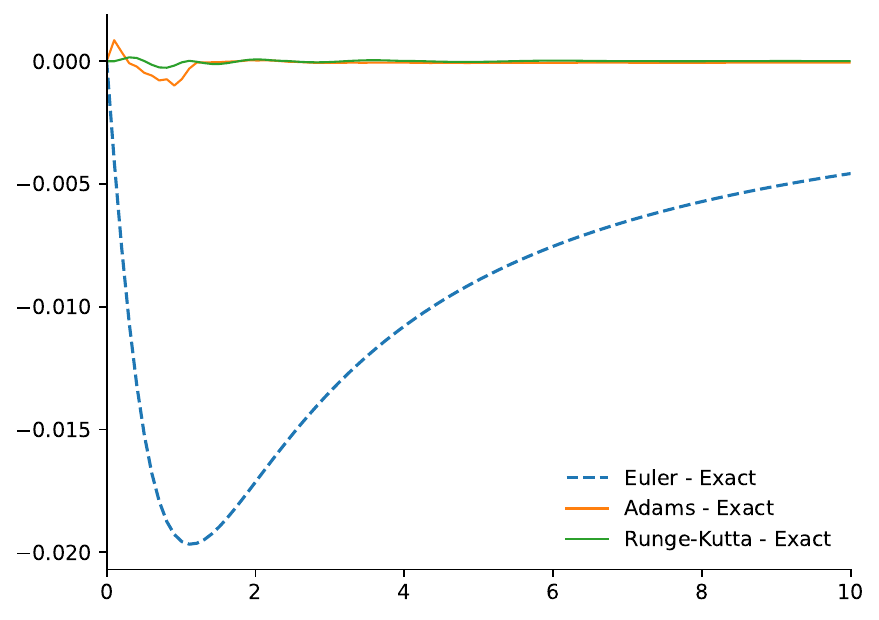}
\caption{The figure on the left is a comparison between the various numerical methods (Euler, Adam, and Runge-Kutta) and the exact solution to equation (\ref{eqn:ode}).  The figure on the right is the difference between various numerical methods and the exact solution.}
\label{fig4}
\end{figure}
Analytically solving the majority of ordinary differential equations, especially in physical contexts, is often challenging or even impossible. However, numerous numerical methods are available for effectively addressing and solving these types of equations. The focus of this work will be on three commonly applied methods of Euler, Adam, and Runge-Kutta.
\subsection{Euler's Method}
 Euler's method serves as an approximation to the solution with initial conditions. While this method may not be the most accurate approximation for practical situations, it stands out as the most straightforward approach for numerical calculation purposes.   Rewriting equation (\ref{eqn:kappa_ode}), 
\begin{equation*}
    \frac{df(x)}{dx}=G(x,f) \nonumber
\end{equation*}
where the initial condition is $f(x_{0})=f_{0}$, the idea for using Euler's Method is to approximate the tangent line of the curve at equally spaced intervals $(x_{i},f(x_{i}))$. The general equation of the tangent line is, 
\begin{equation*}
    f=f_{0}+G(x_{0},f_{0})(x-x_{0}) 
\end{equation*}
By choosing points $(x_{1},f(x_{1}))$, a better approximation can be obtained as $x_{1}$ gets closer to $x_{0}$. As a result, the approximation can further be refined as the distance between $x_{n+1}$ and $x_{n}$ gets smaller, which in turn, allows for an approximation of $f(n)$. This refinement feeds back into even subsequent approximations for $f(n+1)$.

This is written as,
\begin{equation*}
    f_{n+1}=f_{n}+G(x_{n},f_{n})h 
\end{equation*}
where $h=x_{n+1}-x_{n}$ is the step size. By iterating through, the approximate numerical solution to the differential equation can be determined.  

Euler's Method was applied to the $\kappa$-deformed decay differential equation (\ref{eqn:kappa_ode}) and the results and numerical errors along with the analytic solution and comparison to other numerical methods can be seen in figure \ref{fig4}.

\subsection{Adam's Method}
Adam's Method is a multi-step algorithm to find approximate solutions to 
\begin{equation*}
    \frac{df(x)}{dx}=G(x,f). \nonumber
\end{equation*}
This method is attractive due to the fact that it is typically more accurate than Euler's Method, and there is a gain in efficiency requiring only the use of the previous two approximations.  Using the approximation
\begin{equation*}
f_{n+1} = f_n+1 +\frac{3}{2}hG(x_{n+1},f_{n+1})-\frac{1}{2}f(x_{n},y_{n})
\end{equation*}
where $h=x_{n+1}-x_{n}$ is the step size.

Adams's Method was applied to the $\kappa$-deformed decay differential equation (\ref{eqn:kappa_ode}) and the results and numerical errors along with the analytic solution and comparison to other numerical methods can be seen in figure \ref{fig4}.

\subsection{Runge-Kutta Method}
The Runge-Kutta method is the most widely used technique for the numerical solution of general first-order differential equations and stands out as the most accurate among the numerical methods discussed here. It is employed to generate a highly accurate higher-order numerical method without the need for calculating higher-order derivatives. The general form of the Runge-Kutta method is,
\begin{equation}
    f_{n+1} = f_{n}+h\sum_{j=1}^{v}b_{j}K_{j} \nonumber
\end{equation}
where the fourth-order Runge-Kutta method, which is given as, 
\begin{equation}
    f_{n+1} = f_{n}+\frac{h}{6}{(K_{1}+2K_{2}+2K_{3}+K_{4})}  \nonumber
\end{equation}
where $h=x_{n+1}-x_{n}$ is the step size and the slope estimates of the weighted average are given as
\begin{eqnarray}
    K_{1} &=& f{(x_{n},f_{n})}, \nonumber\\
    K_2 &=& f(x_n+\frac{1}{2}h,f_n+\frac{1}{2}K_1), \nonumber\\
    K_3 &=& f(x_n+\frac{1}{2}h,f_n+\frac{1}{2}K_2),\nonumber
\end{eqnarray}
and
\begin{eqnarray}
    K_4 &=& f(x_n+h,f_n+K_3). \nonumber
\end{eqnarray}

Runge-Kutta's Method was applied to the $\kappa$-deformed decay differential equation (\ref{eqn:kappa_ode}) and the results and numerical errors along with the analytic solution and comparison to other numerical methods can be seen in figure \ref{fig4}.

\section{Conclusions}
The objective of this study was to apply elementary methods to solving the deformed $\kappa$-exponential decay differential equation in order to gain experience with deformed statistics, in particular the $\kappa$-deformed exponential at the undergraduate level.  The application of these techniques to equation (\ref{eqn:kappa_ode}) has not been found in the literature except for the method of separation of variables which was applied to 
\begin{equation*}
    \frac{df(x)}{dx}+\frac{\beta}{\sqrt{1+\kappa^2\beta^2x^2}} f(x)=0
\end{equation*}
with the initial condition that $f(0)=1$ and where $\beta$ is a constant.  This resulted in a solution, 
\begin{equation}
\label{eqn:beta_k_decay}
f(x)= \exp_{\kappa}(-\beta x).
\end{equation}
Any of the elementary methods that were applied in this work could be applied to equation (\ref{eqn:beta_k_decay}).

\begin{figure}[h]
\centering
\includegraphics[width=7cm]{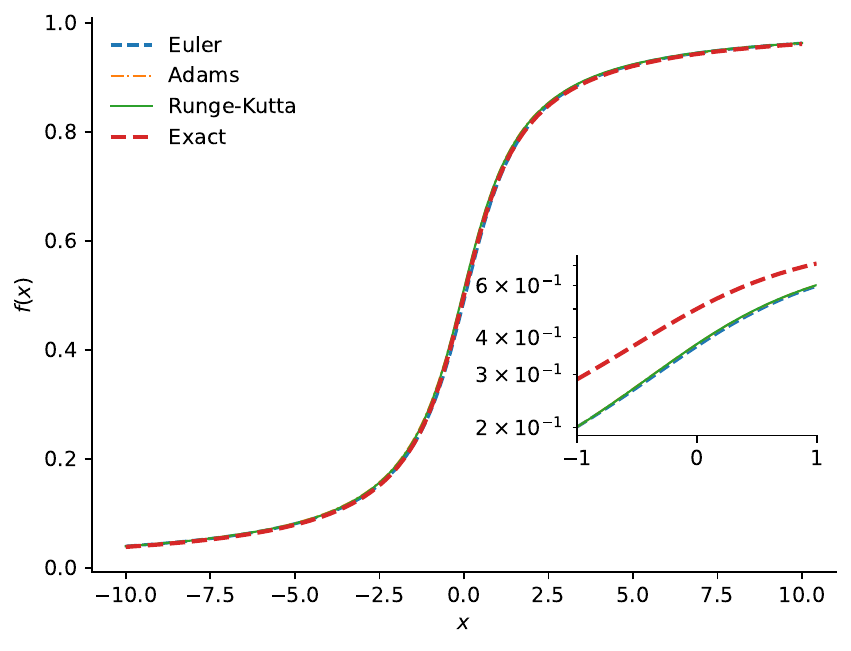}
\includegraphics[width=7cm]{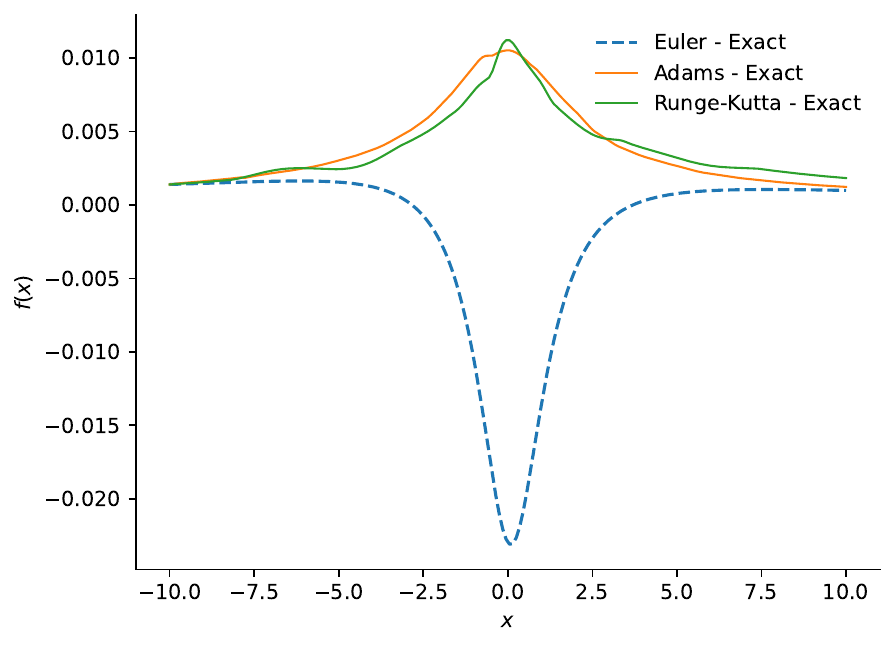}
\caption{The figure on the left is a comparison between the various numerical methods (Euler, Adam, and Runge-Kutta) and the exact solution to the $\kappa$-deformed logistic equation, equation (\ref{eqn:kappa_logistic}).  The figure on the right is the difference between various numerical methods and the exact solution.}
\label{fig5}
\end{figure}

Now that there is experience with the deformed exponential and the deformed decay differential equation, this knowledge can be applied to studying similar differential equations such as: 
\begin{enumerate}
    \item the $\kappa$-logistic differential equation would be an example of the next level of complexity,
    \begin{equation*}
        \frac{df(x)}{dx} = f(x)\Big(1-f(x)\Big) \rightarrow \sqrt{1+\kappa^2 x^2} \frac{df(x)}{dx} = f(x)\Big(1-f(x)\Big)
    \end{equation*}
    where $r$ is the growth rate and $C$ is the carrying capacity.  The solution to the standard logistics differential equation as well as the expected deformed solutions
    \begin{equation}
        f(x)=\frac{1}{1+\exp(-x)} \rightarrow f(x)=\frac{1}{1+\exp_{\kappa}(-x)}
        \label{eqn:kappa_logistic}
    \end{equation} 
    where the solutions along with numerical errors have been plotted in figure \ref{fig5},
    \item analogous differential equations that are important in transport phenomena physics: differential equations such as 
    \begin{enumerate}    
       \item the $\kappa$-advection equation, 
                 \begin{equation*}
                    u\sqrt{1+\kappa^2 x^2}\frac{\partial f(x,t)}{\partial x} = -\frac{\partial f(x,t)}{\partial t}               
                 \end{equation*}
                 where $u$ is the constant velocity associated with the advection, 
       \item $\kappa$-wave equation, 
           \begin{equation*}
                \kappa^2 x \frac{\partial f(x,t)}{\partial x} +(1+\kappa^2 x^2)\frac{\partial^2 f(x,t)}{\partial x^2} = \frac{1}{v^2}\frac{\partial^2 f(x,t)}{\partial t^2}
           \end{equation*}
           where $v$ is the speed of the wave,
       \item $\kappa$-diffusion equation (and other Schrodinger-like equations),
                 \begin{equation*}
                \kappa^2 x \frac{\partial f(x,t)}{\partial x} +(1+\kappa^2 x^2)\frac{\partial^2 f(x,t)}{\partial x^2} = \frac{1}{D}\frac{\partial f(x,t)}{\partial t}
           \end{equation*}
           where $D$ is the diffusion constant.  The $\kappa$-diffusion equation has been examined for the first time by \cite{k_diffusion_eqn1,k_diffusion_eqn2}, and
       \item $\kappa$-Fokker-Planck equation,
                \begin{eqnarray*}
                \kappa^2 x D\frac{\partial f(x,t)}{\partial x} +(1&+\kappa^2 x^2)D\frac{\partial^2 f(x),t}{\partial x^2} \\
                &-\mu\sqrt{1+\kappa^2 x^2}\frac{\partial f(x,t)}{\partial x} = \frac{\partial f(x,t)}{\partial t}
           \end{eqnarray*}
           where $D$ is a diffusion constant and $\mu$ is a drift constant.  The $\kappa$-Fokker-Planck equation has been preliminary studied by \cite{k_fokker_planck1, k_fokker_planck2}.
    \end{enumerate}
    \item fractional differential equations: these differential equations of non-integer order are used to more accurately model complex phenomena.  They utilize fractional derivative operators such as the Caputo derivative.  Preliminary work in this emerging field can be found in \cite{k_frac}, and
    \item expansion of techniques: methods such as Laplace and Fourier transforms \cite {k_fourier1, k_math} have only begun to be investigated and there is still much work to do.  In this work, the Laplace transform was performed using the $\kappa$-numbers.  In the works cited, the authors investigated the $\kappa$-Laplace transform and the $\kappa$-Fourier transform. 
\end{enumerate}

\section*{Acknowledgment}
J. A. Secrest would like to thank Dr. J. M. Conroy at the State University of New York at Fredonia for many enlightening and stimulating conversations about the $\kappa$-formalism. R. Bolle and I. Jarra gratefully acknowledge the financial support of the College of Science and Mathematics at Georgia Southern University.

\end{document}